\documentclass[twocolumn,prb,aps,showpacs,superscriptaddress]{revtex4}
\usepackage{amssymb}

\usepackage{amsmath}
\usepackage{graphicx}
\usepackage{dcolumn}
\usepackage{bm}

\begin{document}

\preprint{}
\title{Theory of electron spin decoherence by interacting nuclear spins in a quantum dot}
\author{Wang Yao}
\affiliation{Department of Physics, University of California San
Diego, La Jolla, California 92093-0319}
\author{Ren-Bao Liu}
\affiliation{Department of Physics, University of California San
Diego, La Jolla, California 92093-0319}
\affiliation{Department of
Physics, The Chinese University of Hong Kong, Shatin, N.T., Hong
Kong, China}
\author{L. J. Sham}
\affiliation{Department of Physics, University of California San
Diego, La Jolla, California 92093-0319}
\date{\today}

\begin{abstract}
We present a quantum solution to the electron spin decoherence by
a nuclear pair-correlation method for the electron-nuclear spin
dynamics under a strong magnetic field and a temperature high for
the nuclear spins but low for the electron. The theory
incorporates the hyperfine interaction, the intrinsic (both direct
and indirect) nuclear interactions, and the extrinsic nuclear
coupling mediated by the hyperfine interaction with the single
electron in question. The last is shown to be important in
free-induction decay (FID) of the single electron spin coherence.
The spin echo eliminates the hyperfine-mediated decoherence but
only reduces the decoherence by the intrinsic nuclear
interactions. Thus, the decoherence times for single spin FID and
ensemble spin echo are significantly different. The decoherence is
explained in terms of quantum entanglement, which involves more
than the spectral diffusion.
\end{abstract}


 \maketitle

\section{Introduction}
\label{sec-intro}

Irreversible processes of a microscopic system in contact with a
macroscopic system are central to nanoscience and to quantum
information science. A canonical example is the spin decoherence
and relaxation of an electron localized by an impurity, an
electrical gate, or a quantum dot in a semiconductor, which have
been extensively studied both in
theory~\cite{Anderson_SpectralDiffusion,Loss_decoherence_nuclei,
Merkulov_decoherence_nuclei,Espin_HF_dipole_1_DaSSarma,
Espin_HF_dipole_3_DaSSarma,shenvi,Espin_HF_1_Loss,
Semenov_SpinT2_phonon,Loss_SpinT2_phonon,Nazarov_Spinflip1,Woods_2002,Espin_HF_Hu}
and in
experiments.~\cite{Fujisawa_2002,1shot_r_Kouwenhoven,abstreiter,Imamoglu_QDSpinPrep,Gurudev,Gammon_T2star,Kouwenhoven_singlet_triplet,T2star_Marcus}
Single dot spin relaxation time $T_1 \gtrsim  0.1 ~$ms has been
measured for different quantum dot systems at low
temperature,~\cite{Fujisawa_2002,1shot_r_Kouwenhoven,abstreiter,Imamoglu_QDSpinPrep}
in good agreement with theoretical
estimates.~\cite{Nazarov_Spinflip1,Woods_2002} The single spin
decoherence time $T_2$ has a lower bound of 10~ns established by
the measurements of the inhomogeneously broadened $T_2^{\ast}$ for
either a spatial ensemble of many dots~\cite{Gurudev} or a time
ensemble of a single
dot.~\cite{Gammon_T2star,T2star_Marcus,Kouwenhoven_singlet_triplet}
Spin dephasing by phonon-scattering in quantum dots is suppressed
at temperature below a few
Kelvins,~\cite{Semenov_SpinT2_phonon,Loss_SpinT2_phonon} leaving
the nuclear spins as the dominant mechanism for electron spin
dephasing. Refs.~\onlinecite{Espin_HF_1_Loss,shenvi} gave,
respectively, a theory and a numerical study of the effects of
off-diagonal electron nuclear hyperfine interaction.
Refs.~\onlinecite{Stamp_spinbath,Espin_HF_dipole_1_DaSSarma,Espin_HF_dipole_3_DaSSarma}
gave treatments of the effects of nuclear dipolar interaction and
calculated the electron spin decoherence time with an ensemble
spin echo.

Our pursuit of a quantum theory of decoherence without the
restriction of a stochastic
theory~\cite{Anderson_SpectralDiffusion,Espin_HF_dipole_1_DaSSarma}
is motivated by the need to control the electron spin decoherence.
The nuclear spins coupled to the electron by the hyperfine
interaction are taken to be the sole source of decoherence as in
Ref.~\onlinecite{Loss_decoherence_nuclei,Espin_HF_1_Loss}.
However, we treat the interaction between nuclear spins which will
be shown to be important in the high magnetic field regime while
the neglect of the nuclear spin interaction in
Ref.~\onlinecite{Loss_decoherence_nuclei,Espin_HF_1_Loss} is valid
in the low field regime.  Our method of solution of the many-spin
problem keys on the evolution of the two flip-flop states of each
pair of nuclear spins and is thus termed the pseudospin method. We
establish conditions for the validity of our method. The method is
simple enough for many applications, including the more advanced
design of pulse control of the electron spin to eliminate the
decoherence effects.\cite{yls2} It also produces a simple physical
picture, which greatly aids the applications. We divide the
interaction between two nuclear spins into two types, intrinsic
and extrinsic, respectively independent and dependent on the
single electron spin state. The intrinsic interaction consists of
the dipole-dipole coupling and  the indirect coupling mediated by
virtual inter-band spin transitions via the hyperfine
interaction.~\cite{indirect_exchange_Bloembergen,indirect_exchange_Anderson,
indirect_exchange_Shulman1,indirect_exchange_Shulman3,indirect_exchange_Sundfors}
The extrinsic nuclear interaction is mediated by virtual spin
flips between each of the two nuclei and the single electron due
to the off-diagonal hyperfine coupling.

The results presented here are (i) a basic solution of the
decoherence dynamics, both for a single electron spin and for an
ensemble of independent electrons; (ii) numerical evaluations for
GaAs dots; and (iii) an analysis of the different time dependence
of coherence for intrinsic and extrinsic nuclear-nuclear
interaction. We will show that the extrinsic hyperfine-mediated
nuclear interaction plays an important role in single spin FID.
The spin echo not only refocuses the dephasing by inhomogeneous
broadening in ensemble dynamics but also eliminates the
decoherence by extrinsic hyperfine-mediated nuclear interaction.
Thus, the decoherence times for single spin FID and spin echo are
significantly different. The usual practice of inferring the
single electron spin dephasing time from ensemble echo measurement
could be problematic. Note that in NMR literature, the difference
in timescale of FID and echoes, when the inhomogeneous effect is
excluded, was first recognized.\cite{pines,ernst}

The electron spin decoherence arises out of quantum entanglement
between the electron spin states $|\pm\rangle$ and the
many-nuclear spin states $|{\mathcal J}\rangle$. When a coherent
electron spin state $C_+|+\rangle +C_-|-\rangle$ is prepared, the
initial state of the whole electron-nuclear system is the product
state, $(C_+|+\rangle +C_-|-\rangle)\otimes |{\mathcal J}\rangle$.
In time $t$, the nuclear states associated with the two electron
spin states diverge, yielding an entangled state of the form,
$C_+|+\rangle\otimes |{\mathcal J}^+(t)\rangle +
C_-|-\rangle\otimes |{\mathcal J}^-(t)\rangle$. The electron spin
coherence is measured by $|\langle {\mathcal J}^-(t)|{\mathcal
J}^+(t)\rangle|$ when the nuclear spin degrees of freedom are
traced out. Our theory consists in a direct attack of the many
nuclear spin dynamics. The assumption of the pure electron spin
decoherence time $T_2$ being much shorter than its longitudinal
spin relaxation time $T_1$ will be shown to lead to a simple
effective Hamiltonian for the whole system of the form
$\sum_{\pm}|\pm\rangle \hat{H}_{\pm} \langle\pm|$ in terms of the
nuclear spin Hamiltonians $\hat{H}_{\pm} $. This  simplifies the
electron spin coherence to the overlap of the two nuclear spin
states, each following the evolution conditioned on one spin state
of the electron,  $|\langle {\mathcal J}|e^{- \hat{H}_-t }e^{i
\hat{H}_+t}|{\mathcal J}\rangle|$.

This entanglement approach to decoherence has an interesting
relation to the antecedents in the decoherence literature. The
square of the decoherence is formally the same as the Loschmidt
echo if $\hat{H}_-$, say, is regarded as  $\hat{H}_+$ with  a
perturbation, which is related to the decoherence of the nuclear
spin system. \cite{zurek03} A model of a single spin coupled to a
transverse field Ising chain is used to study the effect of
quantum phase transition on the decoherence of the Ising
chain.\cite{sun} A key change in the model could make it a study
of the decoherence of the single spin in the Ising spin bath.

We will show that the nuclear spin dynamics is dominated by the
nuclear spin pair flips. The pairs can be treated as independent
of one another and only the two-spin correlations need be taken
into account in the interacting nuclear spin dynamics. On a time
scale small compared with the inverse nuclear couplings but ample
for the electron spin decoherence, the number of pair-flip
excitations are small compared with the number of nuclear spin
pairs available for spin flip, which come from the randomization
of the nuclear spin directions at a temperature higher than the
nuclear spin temperature, i.e., 10~mK $\lesssim T \lesssim$ 1~K
(still low enough to avoid the effects of the electron-phonon
scattering). The cluster expansion by Witzel \textit{et
al.}~\cite{Espin_HF_dipole_3_DaSSarma} yields an equivalent
pair-correlation approximation. We will establish the pseudo-spin
model in which the elementary excitations by independent
pair-flips are just rotations of non-interacting 1/2 spins.

Numerical evaluations for a GaAs dot then require no further
approximation. The electron spin decoherence is calculated for a
range of magnetic field strength (1--40~T) and various dot sizes.
The pseudo-spin rotation yields the following analysis of the
results. The $e^{-t^n}$ short-time behavior obeys $n=2$ or $4$
depending on the dominance of, respectively, the extrinsic and the
intrinsic nuclear-nuclear interaction. In the long time limit, the
crossover to the exponential decay ($n=1$) indicates the onset of
the Markovian kinetics.

The main body of the text gives a succinct account of the key
points of the theory and the results of the computation.
Section~\ref{sec-2} defines the single electron system coupled to
a bath of interacting nuclear spins. Section~\ref{sec-decoh}
defines the electron spin coherence and formulates the quantum
theory of its evolution. Section~\ref{sec-pss} describes the
pseudo-spin solution. Section~ \ref{sec-results} gives an
evaluation of the decoherence for a GaAs quantum dot.
Section~\ref{sec-summary} serves as a brief summary of the main
results. So as not to interrupt the flow of the essence of simple
exposition of our decoherence theory, further details of the
theory are grouped in the Appendices.

\section{Single electron in interacting nuclear spin bath} \label{sec-2}

The system consists of an electron with spin vector $\hat{{\mathbf
S}}_{\rm e}$ and $N$ nuclear spins, $\hat{{\mathbf J}}_{n}$, with
Zeeman energies $\Omega_{\rm e}$ and $\omega_{n}$ under a magnetic
field $B_{\rm ext}$, respectively, where $n$ denotes both
positions and isotope types (e.g. $^{75}$As, $^{69}$Ga and
$^{71}$Ga in GaAs). The Hamiltonian of this system is described in
Appendix~\ref{app-H}. The interaction can be separated as
``diagonal'' terms which involve only the spin vector components
along the field ($z$) direction and ``off-diagonal'' terms which
involve spin flips (see Appendix~\ref{Append_transformation}).
Because the electron Zeeman energy is much larger than the
strength of the hyperfine interaction, the off-diagonal term is
eliminated by a standard canonical transformation, with the
second-order correction left as the hyperfine-mediated nuclear
interaction. For the same reason, the off-diagonal part of the
nuclear interaction contributes only when the terms conserve the
Zeeman energies (so-called secular terms in the NMR terminology).
Hence, the non-secular terms are negligible. The total reduced
Hamiltonian is obtained from the transformation in
Appendix~\ref{Append_transformation}, for the limit of long
longitudinal relaxation time ($T_1 \rightarrow \infty$),
\begin{equation}
 \hat{H}_{\rm red}=\hat{H}_\text{e}+\hat{H}_\text{N}+\sum_{\pm}|\pm\rangle \hat{H}_{\pm} \langle\pm|,
 \label{eq-effH}
\end{equation}
with $\hat{H}_{\rm e}=\Omega_{\rm e} \hat{S}^z_{\rm e}$,
$\hat{H}_{\rm N}=\omega_{n}\hat{J}^z_{n}$, and the interaction
terms,
\begin{equation}
\hat{H}_{\pm}=\pm \hat{H}_A+\hat{H}_B+\hat{H}_D\pm \hat{H}_E, \label{eq-Hpm}
\end{equation}
given by,
\begin{subequations} \begin{eqnarray}
\hat{H}_A &=&
  {\sum_{n\ne m}}'\frac{a_n a_m}{4  \Omega _{\rm e}}
  \hat{J}_n^{+}\hat{J}_m^{-}\equiv {\sum_{n \ne m}}' A_{n,m} \hat{J}_n^{+}\hat{J}_m^{-}, \ \ \ \   \label{HA} \\
\hat{H}_B &=&{\sum_{n \ne  m}}' B_{n,m} \hat{J}_n^{+}\hat{J}_m^{-}
\label{HB}  \\
\hat{H}_D &=&{\sum_{n < m}} D_{n,m} \hat{J}_n^{z}\hat{J}_m^{z}
\label{HD}  \\
\hat{H}_E &=&
 \sum_{n} \frac{a_n}{2}\, \hat{J}_n^{z}\equiv \sum_{n}E_n \hat{J}_n^{z},   \label{HE}
\end{eqnarray}  \label{Hamiltonian}
\end{subequations}
where $|\pm\rangle$ are the eigenstates of $\hat{S}^z_{\rm e}$,
the summation with a prime runs over only the homo-nuclear pairs,
the subscript $A$ denotes the extrinsic hyperfine mediated
interaction, $B$ the off-diagonal part of the intrinsic
interaction, $D$ the diagonal part of the intrinsic interaction,
and $E$ the diagonal part of the contact electron-nuclear
hyperfine interaction. The hyperfine energy,~\cite{Paget}
determined by the electron wavefunction, has a typical energy
scale $E_n\sim a_n \sim 10^6$~s$^{-1}$ for a dot with about $10^6$
nuclei. The sum, $\mathcal{A} \equiv \sum_n a_n$, is the {\it
hyperfine constant} depending only on the material. The intrinsic
nuclear spin-spin interaction has the near-neighbor coupling
$B_{n,m}\sim {D}_{n,m} \sim 10^2$~s$^{-1}$. The hyperfine mediated
interaction, which is unrestricted in range and associated with
opposite signs for opposite electron spin states, has an energy
scale dependent on the field strength, $A_{n,m}\sim
1$--$10$~s$^{-1}$ for field $\sim 40$--$1$~T.

\section{Decoherence Theory}  \label{sec-decoh}

The electron-nuclear spin system is assumed to be initially prepared in a product state with the nuclear spins in a thermal state with temperature $T$, described by the density matrix
\begin{eqnarray}
\hat{\rho}(0)=\hat{\rho}^{\rm e}(0)\otimes \hat{\rho}^{\rm N}.
\end{eqnarray}
The time evolution of the reduced density matrix of the electron spin,
\begin{eqnarray}
\hat{\rho}^{\rm e}(t)={\rm Tr}_{\rm N} \hat{\rho}(t),
\end{eqnarray}
obtained by tracing over the nuclear spins, may be expressed in the form,
\begin{eqnarray}
\rho^{\rm e}_{\mu,\nu}(t)=\sum_{\mu',\nu'}{\mathcal L}_{\mu,\nu; \mu',\nu'}(t) \rho^{\rm e}_{\mu',\nu'}(0)
\end{eqnarray}
where $\rho^{\rm e}_{\mu,\nu}\equiv \langle \mu|\rho^{\rm e}|\nu\rangle$, and
$|\mu\rangle$, $|\nu\rangle\in\{|+\rangle,\ |-\rangle\}$. The superoperator or
 correlation function ${\mathcal L}_{\mu,\nu; \mu',\nu'}$ can be expressed in terms of the evolution operator and contains the information on the electron spin relaxation and decoherence.

The Hamiltonians of Eq.~(\ref{eq-effH}) for the $T_1 \rightarrow
\infty$ limit conserves the electron $\hat{S}^e_z$ quantum number:
$[\hat{H},\hat{S}_z^e]=0$. Hence, the correlation function has
following properties,
\begin{subequations}
\begin{eqnarray}
{\mathcal L}_{\mu,\nu; \mu',\nu'}(t)&=& {\mathcal L}_{\mu,\nu}(t)\delta_{\mu,\mu'}\delta_{\nu,\nu'}, \\
{\mathcal L}_{\mu,\mu}(t)&=&1, \\
{\mathcal L}_{+,-}(t)&=& {\mathcal L}^*_{-,+}(t),
\end{eqnarray}
\end{subequations}
and the specific expression for the free-induction decay,
\begin{equation}
{\mathcal L}_{+,-}(t) =  e^{-i\Omega_e t}{\rm Tr}_{\rm N}
\left[\hat{\rho}^N e^{+i\hat{H}_- t}e^{-i\hat{H}_+ t}\right],
\end{equation}
which can straightforwardly extended to dynamics under pulse
control.

The ensemble of nuclear spins, at temperature $T \gtrsim \omega_n
\gg A_{n,m}, B_{n,m}, D_{n,m}, E_n$, may be approximated by the
density matrix,
\begin{eqnarray}
\hat{\rho}^{\rm N} \approx e^{-\hat{H}_{\rm N}/T} = \sum_{\mathcal
J}P_{\mathcal J}|{\mathcal J}\rangle\langle {\mathcal J}|,
\label{thermal}
\end{eqnarray}
where $|{\mathcal J}\rangle\equiv \bigotimes_n |j_n\rangle$, $j_n$
being the quantum number of nuclear spin $n$ in the magnetic field
direction. $P_{\mathcal J}$ is the thermal distribution factor.
The correlation function ${\mathcal L}_{+,-}(t)$ can then be
generally expressed as,
\begin{equation}
{\mathcal L}_{+,-}(t) = \sum_{\mathcal J} P_{\mathcal J}  e^{-i
\phi_{\mathcal J}(t)} \left|\langle \mathcal{J}^- (t) |
\mathcal{J}^+ (t) \rangle\right|
\end{equation}
In FID, $| \mathcal {J}^{\pm} (t) \rangle = e^{-i \hat{H}^{\pm} t}
| \mathcal{J} \rangle$ and $\phi_{\mathcal J}(t)=(\Omega_e +
\mathcal{E}_{\mathcal J}) t$ where ${\mathcal
E}_\mathcal{J}=\sum_n j_n a_n$ is the contribution to the electron
zeeman splitting from the Overhauser field in the nuclear
configuration $| \mathcal{J} \rangle$. With a $\pi$ pulse to flip
the electron spin at time $\tau$, we have $| \mathcal {J}^{\pm}
(t>\tau) \rangle =e^{-i \hat{H}^{\mp} (t-\tau)} e^{-i
\hat{H}^{\pm} \tau} | \mathcal{J} \rangle$ and $\phi_{\mathcal
J}(t>\tau)=(\Omega_e + \mathcal{E}_{\mathcal J}) (2\tau-t)$.

An important finding is that the coherence for the ensemble
dynamics takes the factored form,
\begin{equation}
{\mathcal L}_{+,-}(t) = {\mathcal L}^{\rm s}_{+,-}(t)\times {\mathcal L}^{(0)}_{+,-}(t), \label{e_dm_4}
\end{equation}
where
\begin{subequations}
\begin{eqnarray}
{\mathcal L}^{\rm s}_{+,-}(t) &= & \left|\langle {\mathcal J}^- (t) | {\mathcal J}^+ (t) \rangle\right|,
\label{eq-losch} \\
{\mathcal L}^{(0)}_{+,-}(t)  &= & \sum_{\mathcal J} P_{\mathcal J}
e^{-i\phi_{\mathcal J}(t)}. \label{eq-sd}
\end{eqnarray}
\end{subequations}
${\mathcal L}^s_{+,-}(t)$ characterize the electron coherence
evolution in the {\it single-system dynamics} with the nuclear
bath begins on a typical pure initial state $|{\mathcal J}\rangle
\equiv \bigotimes _n |j_n\rangle$. The single-system dynamics is
solely determined by the spectrum of the elementary excitations
from the initial state $|{\mathcal J}\rangle$, i.e., the nuclear
spin pair-flips driven by $\hat{H}_A$ and $\hat{H}_B$ in
Eqs.~(\ref{HA},\ref{HB}). For a sufficiently large number $N$ of
the nuclear spins, the spectrum is independent of the initial
state with an error of the order of $1/\sqrt{N}$. Therefore the
coherence factor ${\mathcal L}^s_{+,-}(t)$ can be pulled out of
the sum over ${\mathcal J}$ which results in the factorized form
as in Eq.~(\ref{e_dm_4}). The independence of the nuclear pair
excitation spectrum on the initial state has been numerically
verified with a number of different initial states in a GaAs
quantum dot with $N = 10^6$. The independence of ${\mathcal
L}^s_{+,-}(t)$ on the initial nuclear state $|\mathcal{J}\rangle$
makes the measurement of single-system coherence possible in
principle, e.g., in single spin measurement with pre-determination
or post-selection of local Overhauser field through projective
measurement.\cite{Espin_HF_1_Loss}

The ensemble effect resides entirely in the factor ${\mathcal
L}^{(0)}_{+,-}(t)$, which may be read as the inhomogeneous
broadening of the local field ${\mathcal E}_\mathcal{J}$ with
distribution function $P_{\mathcal J}$. The inhomogeneous factor
(i.e., the distribution of the hyperfine energy ${\mathcal
E}_\mathcal{J}$) dominates the free induction decay in the
ensemble dynamics in the form of ${\mathcal L}^{(0)}_{+,-}(t)=
e^{-i\Omega_{\rm e} t-(t/T_2^*)^2}$, with the dephasing time
$T_2^* \sim \sqrt{N} \mathcal{A}^{-1}\sim 10$~ns as
measured.~\cite{Gammon_T2star,T2star_Marcus,Kouwenhoven_singlet_triplet,Gurudev}
To single out the dynamical decoherence time from the $T_2^*$,
spin echo pulses can be applied to eliminate the effects of the
static fluctuations of the local field. After a $\pi$-pulse
applied at $\tau$, the inhomogeneous broadening part of the
correlation function ${\mathcal L}^{(0)}_{+,-}(t)=1$ for
$t=2\tau$. The ensemble coherence peak at $2\tau$ is known as spin
echo.\cite{Hahn} The spin echo profile, i.e., the echo magnitude
${\mathcal L}_{+,-}(2\tau) = {\mathcal L}^{\rm s}_{+,-}(2\tau)$
plotted as a function of the echo delay time $2\tau$, reveals the
dynamical processes that leads to decoherence.

\section{The pseudo-spin solution} \label{sec-pss}

The solution to the single-system evolution
$|\mathcal{J}^{\pm}(t)\rangle$ relies on the pair-correlation
approximation explained here with more details in
Appendix~\ref{A_solution} and justified in
Appendix~\ref{A_validity}. Within a time $t$ much smaller than the
inverse nuclear interaction strength, the total number of
pair-flip excitations $N_{\rm flip}$ is much smaller than the
number of nuclei $N$. The probability of having pair-flips
correlated is estimated in Appendix~\ref{SS_PCA} to be $P_{\rm
corr}\sim 1-e^{-q N_{\rm flip}^2/N}$ ($q$ being the number of
homo-nuclear nearest neighbors), which, as also shown by \textit{a
posteriori} numerical check, is well bounded by $\sim 1 \%$ in the
worst scenario studied in this paper. Thus, the pair-flips as
elementary excitations from the initial state can be treated as
independent of each other, with a relative error $\epsilon
\lesssim P_{\rm corr}$. Then the single-system dynamics
$|\mathcal{J}^{\pm} (t)\rangle$ can be described by the excitation
of pair-correlations as non-interacting quasi-particles from the
``vacuum'' state $|\mathcal{J}\rangle$, driven by the
``low-energy'' effective Hamiltonian,
\begin{equation}
\hat{H}^\pm_{\mathcal J} = \sum_k \hat{\mathcal H}^{\pm}_k\equiv
\sum_k {\mathbf h}_k^{\pm} \cdot \hat{\boldsymbol \sigma}_k/2,
\label{pseudospin_Hamil}
\end{equation}
which has been written in such a way that the pair-correlations
are interpreted as $1/2$-pseudo-spins, represented by the Pauli
matrix $\hat{\boldsymbol \sigma}_k$, with $k$ labelling all
possible pair-flips, defined in more details in
Appendix~\ref{Asub_PseudoSpin}. The time evolution from the
initial state $| \mathcal{J} \rangle $ can be viewed as the
rotation of the pseudo-spins, initially all polarized along the
$+z$ pseudo-axis: $\bigotimes_k |\uparrow_k\rangle $, under the
effective pseudo-magnetic field,
\begin{equation}
 {\mathbf h}^{\pm}_k \equiv (\pm
2A_k+2B_k,0, D_k \pm E_k),
\end{equation}
where, for the electron spin state $|\pm\rangle$, $\pm A_k$ and
$B_k$ are the pair-flip transition amplitudes, defined in
Eqs.~(\ref{elementaryexcitation}), contributed by the hyperfine
mediated coupling $\hat{H}_A$ and the intrinsic coupling
$\hat{H}_B$, respectively, and $D_k$ and $\pm E_k$ are the energy
cost of the pair-flip contributed by the diagonal nuclear coupling
$\hat{H}_D$ and the hyperfine interaction $\hat{H}_E$,
respectively. The decoherence then can be analytically derived as
\begin{eqnarray}
{\mathcal L}^s_{+,-}(t) = \prod_k\left|\langle\psi^-_k|\psi^+_k\rangle\right|
\approx\prod_k e^{-\delta_k^2/2},
\end{eqnarray}
where $|\psi^{\pm}_k\rangle$ are the conjugate pseudo-spin states.
In FID, $|\psi^{\pm}_k (t) \rangle \equiv e^{-i\hat{\mathcal
H}_k^{\pm} t}|\uparrow_k\rangle$; while with a $\pi$ pulse to flip
the electron at $t=\tau$, $|\psi^{\pm}_k (t
> \tau) \rangle \equiv e^{-i\hat{\mathcal H}_k^{\mp} (t-\tau)}
e^{-i\hat{\mathcal H}_k^{\pm} \tau}|\uparrow_k\rangle$.
$\delta_k^2\equiv
1-\left|\langle\psi^-_k|\psi^+_k\rangle\right|^2$ possesses a
simple geometrical interpretation: the squared distance between
the two conjugate pseudo-spins on the Bloch sphere, which
quantifies the entanglement between the electron spin and the
pseudo-spins.

\section{Results of decoherence and spin echo for a quantum dot} \label{sec-results}

\begin{figure}[t]
\includegraphics[width=8cm, height=4.5cm, bb=10 271 589 593,
clip=true]{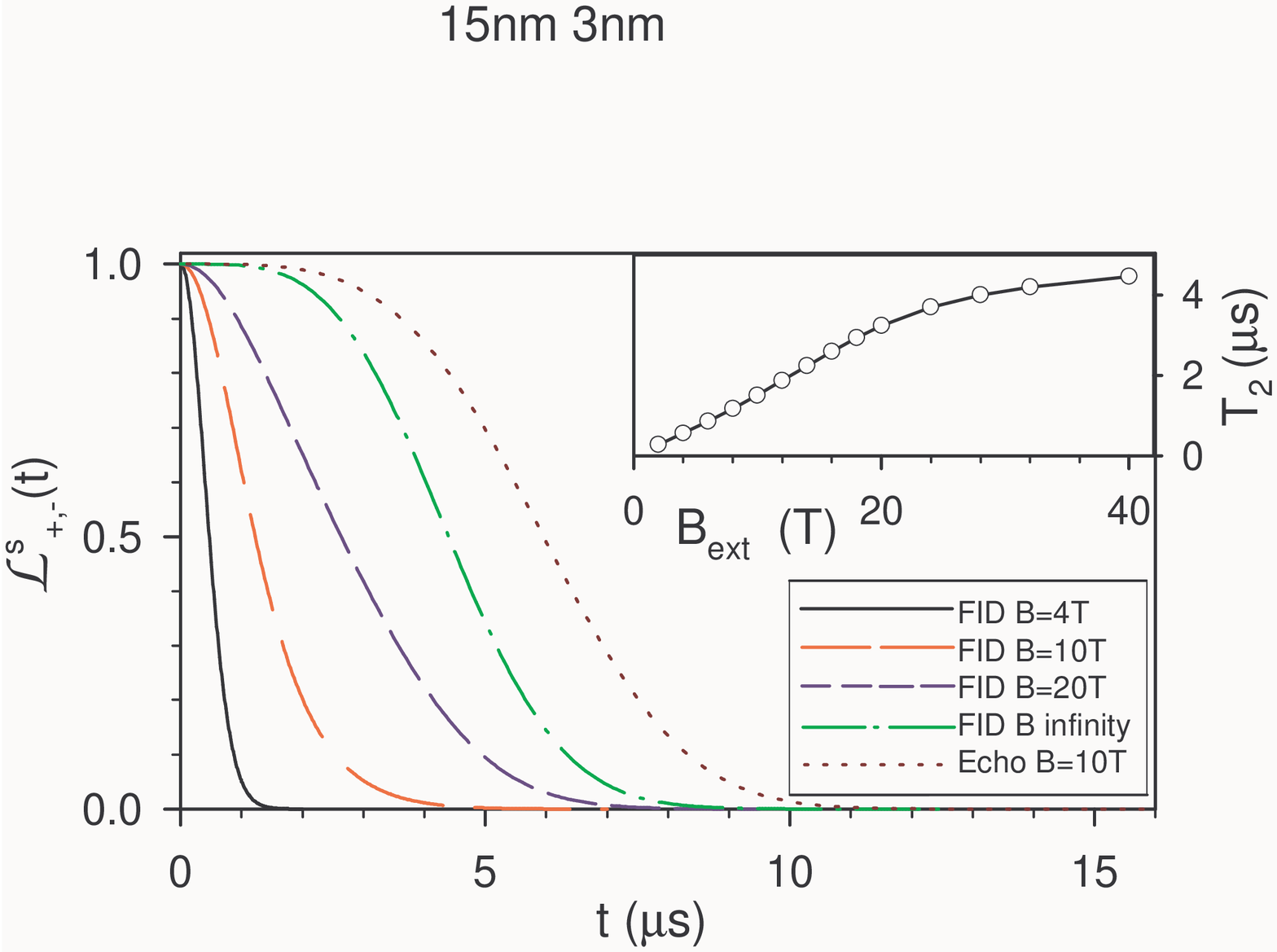} \caption{(Color online) Single-system FID for
a dot with $L_{[001]}=2.8$~nm and $r_0=15$~nm under various field
strengths. The spin echo profile as a function of $t=2\tau$ for
$B_{\rm ext}= 10$~T is also plotted for comparison. The insets
shows the field dependence of the FID decoherence time.}
\label{FID_all}
\end{figure}

In this section, we analyze the results of decoherence under free
induction and spin echo conditions for a GaAs quantum dot. In
numerical evaluations, the GaAs dot is assumed to have a hard-wall
confinement in the growth direction $[001]$ with thickness
$L_{[001]}$ and a parabolic confinement with Fock-Darwin radius
$r_0$ in the lateral directions. The external magnetic field is
applied along the $[110]$ direction. For the indirect intrinsic
nuclear interaction, we consider only the exchange part.
\cite{indirect_exchange_Sundfors} The $g$ factor of the
electron\cite{Gurudev} is taken as $-0.13$. The initial state
$|\mathcal{J}\rangle$ is generated by randomly setting each
nuclear spin according to a ``high-temperature'' Boltzmann
distribution ($P_{\mathcal J}={\rm constant}$).

\begin{figure}[t]
\includegraphics[width=8cm, height=4.7cm, bb=54 150 577 468,
clip=true]{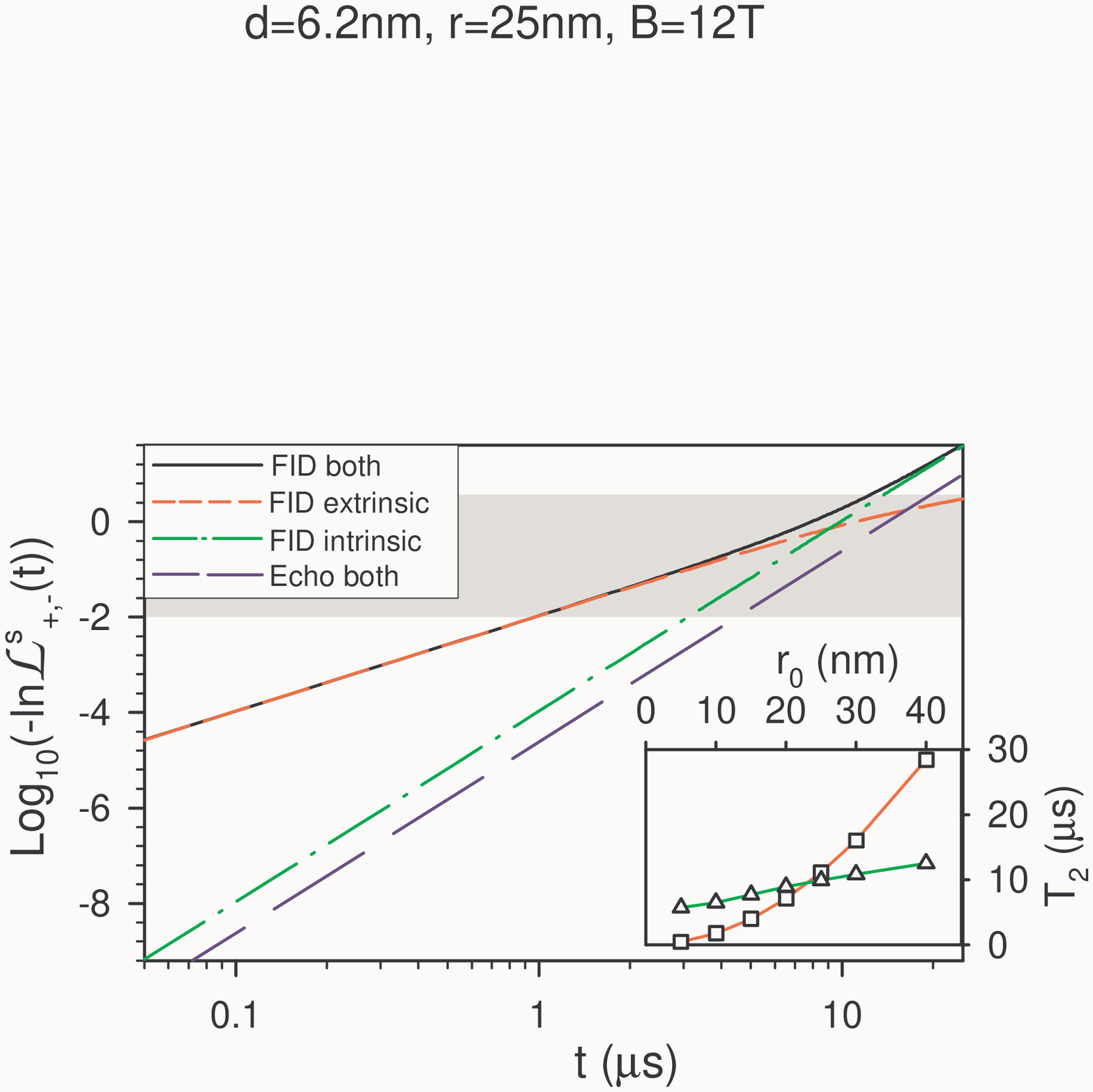} \caption{(Color online) Separated
contributions to the single-system FID by the extrinsic
hyperfine-mediated nuclear coupling, the intrinsic nuclear
interaction, and both, for a dot with $L_{[001]}=6.2$~nm and
$r_0=25$~nm at $B_{\rm ext}=12$~T. The spin echo profile with both
mechanisms is shown in comparison. The inset shows the FID
decoherence times resulting from only the hyperfine-mediated
interaction (square symbol) or only the intrinsic interaction
(triangle symbol) as functions of $r_0$.} \label{FID_sep12}
\end{figure}

\begin{figure}[t]
\includegraphics[width=7cm, height=4.5cm, bb=129 212 558 477, clip=true]{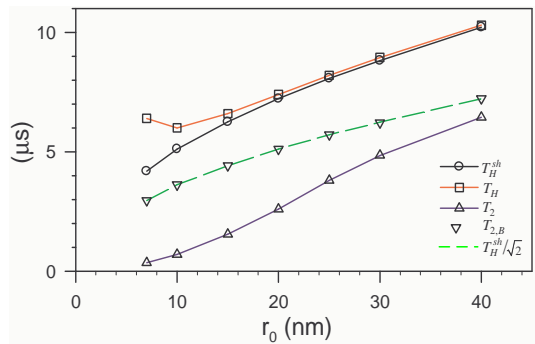}
\caption{(Color online) Dot-size dependence of the decoherence
times (see text), $T_{\rm H}^{\rm sh}$ as circles, $T_{\rm H}$
squares, $T_{2}$ triangles, $T_{2,B}$ inverted triangles. $T_{\rm
H}^{\rm sh}/ \sqrt{2}$ is plotted (dashed blue line) for compare
with $T_{2,B}$. The dot thickness $L_{[001]}=2.8$~nm and $B_{\rm
ext}=10$T. } \label{times}
\end{figure}

Fig.~\ref{FID_all} shows the FID in single-system dynamics for a
typical dot under various field strengths $B_{\rm ext}$. The inset
of Fig.~\ref{FID_all} shows the field dependence of decoherence
time $T_2$ which is defined as the time when the FID signal is
$1/e$ of its initial value. The strong field dependence of $T_2$
demonstrate the significance of the extrinsic hyperfine mediated
nuclear coupling until it is suppressed by a very strong field
($\sim 20$~T). As shown in Fig.~\ref{FID_all}, the FID signals
have significantly different decoherence times from the spin echo
signals (See also Fig.~\ref{times}). In Fig.~\ref{FID_sep12}, we
separate artificially the contributions from the extrinsic
hyperfine-mediated and the intrinsic nuclear interactions and show
their different dependence on time and dot-size. The spin echo
profile is also plotted for comparison.

A couple of justified simplifications can provide an understanding
of the effects of various mechanisms on the spin decoherence.
First, the energy cost by the diagonal nuclear coupling ($D_k$)
can be neglected as it is by three orders of magnitude smaller
than that by hyperfine interaction ($E_k$). Second, for
near-neighbor pair-flips, the intrinsic nuclear interaction is
much stronger than the hyperfine mediated one for the field
strength under consideration. Third, for non-local pair-flips, the
intrinsic interaction is negligible due to its finite-range
characteristic. Thus we can separate the flip-pairs into one
subset, $k \in K_A$, which contains $O(N^2)$ non-local flip-pairs,
driven by the effective pseudo-magnetic field ${\mathbf
h}^{\pm}_k\approx \left(\pm 2A_k,0,\pm E_k\right)$ and a second
subset, $k \in K_B$, which contains $O(N)$ near-neighbor
flip-pairs, driven by ${\mathbf h}^{\pm}_k\approx \left(2B_k,0,\pm
E_k\right)$. The conjugate pseudo-spins will precess along
opposite directions in the non-local subset $K_A$, and
symmetrically with respect to the $y$-$z$ plane in the
near-neighbor subset $K_B$. The decoherence can be readily grouped
by the two different mechanisms as
\begin{eqnarray}
{\mathcal L}^{\rm s}_{+,-}\cong \prod_{k \in
K_B}e^{-\frac{t^4}{2}E_k^2 B^2_k
 {\rm sinc}^4\frac{h_k t }{2}}
 \prod_{k\in K_A}e^{-2 t^2 A_k^2 {\rm sinc}^2 (h_k t)},
 \label{separation}
\end{eqnarray}
where $h_k = | {\mathbf h}^{\pm}_k|$ and ${\rm sinc}(x)\equiv
\sin(x)/x$. We can see that the extrinsic hyperfine-mediated and
the intrinsic couplings lead to the $e^{-(t/T_{2,A})^2}$ and the
$e^{-(t/T_{2,B})^4}$ behavior in time shorter than the inverse
pair-flip energy cost (which corresponds to the width of the
excitation spectrum),
\begin{equation}
T_{2,B}\approx b^{-1/2} \mathcal{A}^{-1/2} N^{1/4}; ~ T_{2,A}\approx
\Omega_e \mathcal{A}^{-2} N \label{size_field_depend}
\end{equation}
where $b$ is the typical value of near neighbor intrinsic nuclear
coupling strength $B_k$ and $\mathcal{A} \equiv \sum_n a_n$ is the
hyperfine constant. Eq.~(\ref{size_field_depend}) explains the
dot-size dependence of the two mechanisms in inset of
Fig.~\ref{FID_sep12}. When the two mechanisms are comparable, the
single-system FID begins with $e^{-t^2}$ behavior and may cross
over towards $e^{-t^4}$ decay as time increases, and this is
actually observed in Fig.~\ref{FID_sep12}. As the time grows
beyond the inverse excitation spectrum width, the quantum kinetics
becomes a stochastic Markovian process by building up the
energy-conserving Fermi-Golden rule as indicated by the sinc
function in Eq.~(\ref{separation}).

\begin{figure}[t]
\includegraphics[width=7.5cm, height=4.5cm, bb=45 140 534 450, clip=true]{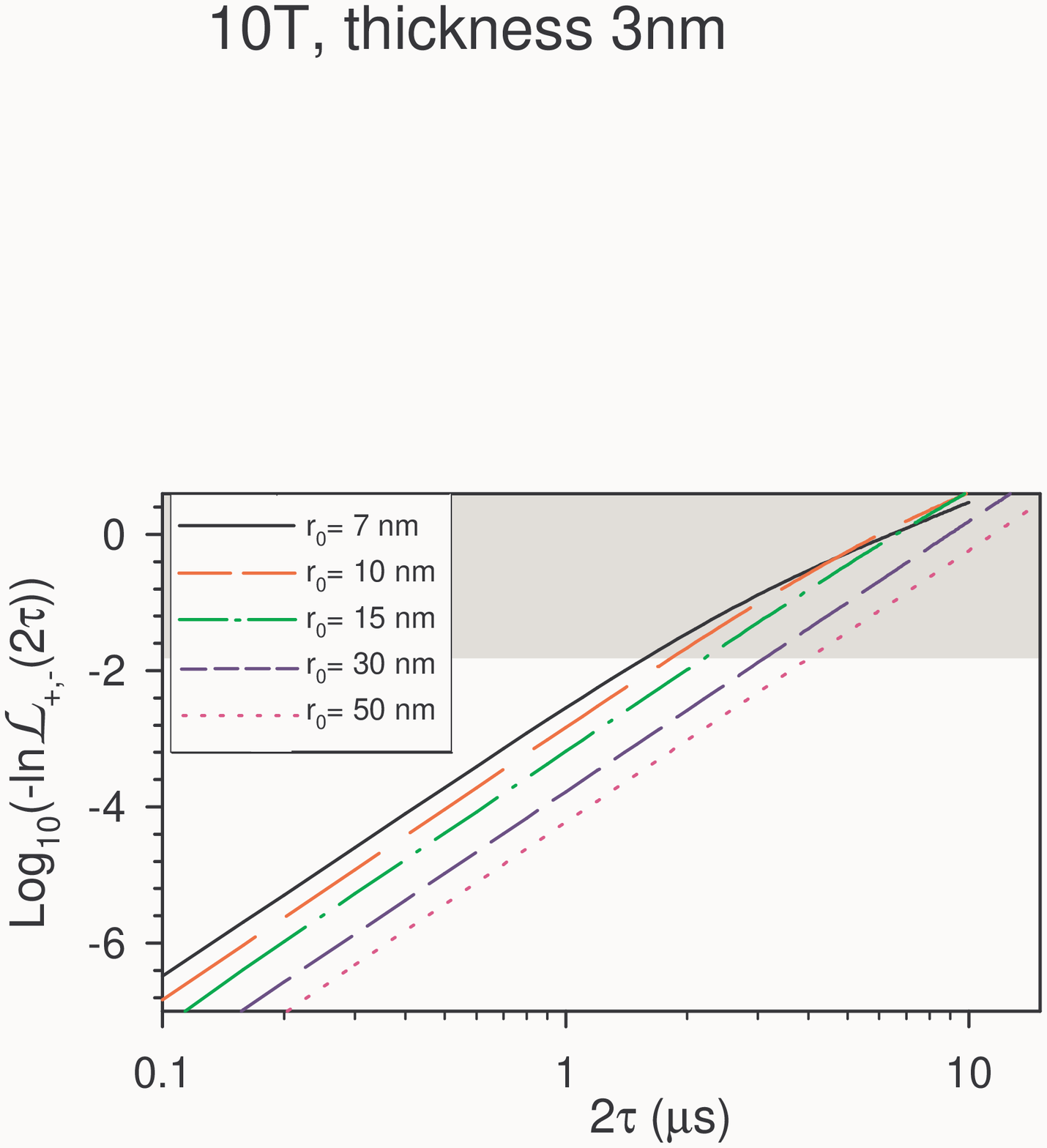}
\caption{(Color online) Spin echo signal for dots of
$L_{[001]}=2.8$~nm and various $r_0$ at $B_{\rm ext}=10$T.}
\label{size}
\end{figure}

Fig.~\ref{size} shows the onset of the Markovian process by the
crossover from the $e^{-t^4}$ short-time behavior to the long-time
exponential decay $e^{-t}$.

In ensemble or repeated
dynamics,~\cite{Gurudev,Gammon_T2star,T2star_Marcus} FID will be
dominated by the inhomogeneous broadening. In ESR
experiments,\cite{ESR_silicon_T2,Abe_Si} electron spin flip pulses
can be applied to eliminate the effects of the static fluctuations
of the local field and spin echo profiles are measured. The echo
decay time $T_{\rm H}$ (defined as the echo delay time at which
the spin echo magnitude drops to $1/e$ of the zero delay value) is
generally believed to give a quantitative measure of the
single-system FID time $T_2$, as the direct measurement of the
latter is of considerable difficulty with current experimental
capability. We now show that the echo pulse will also modify the
electron spin decoherence induced by the quantum pair-flip
dynamics, and as a consequence, $T_{\rm H}$ and $T_2$ can be
significantly different. As the electron spin is reversed by the
$\pi$-pulse, the hyperfine-mediated transition amplitude $A_k$ and
the hyperfine energy cost $E_k$ for each pair-flip will change the
sign after the pulse. Thus, the pseudo-spins driven by the
extrinsic hyperfine-mediated nucear coupling (in subset $K_A$)
will reverse their precession after the pulse and return to the
origin at $t=2\tau$, disentangling the electron spin and the
pseudo-spins. So the decoherence driven by the extrinsic
hyperfine-mediated coupling is largely eliminated in the spin-echo
configuration (see Fig.~\ref{FID_sep12}). This phenomenon was
first noted in the numerical simulation of a small system of $\sim
10$ nuclear spins in \cite{shenvi}. For the pseudo-spin driven by
the intrinsic coupling (subset $K_B$), the conjugate pseudo-spins
will switch their precession axis which also reverse the
entanglement to some extent but no full recover can be obtained at
the echo time. Finally, the electron spin coherence at the echo
time can be derived as
\begin{eqnarray}
{\mathcal L}_{+,-}(2\tau) \cong \prod_{k\in K_B} e^{-{2 \tau^4
E_k^2B_k^2} {\rm sinc}^4\left(h_k^B\tau/2\right)} \label{d-echo}.
\end{eqnarray}
Similar to the analysis for single system FID, the spin echo
signal begins with $e^{-(2\tau /T_{\rm H}^{\rm sh})^4}$ short-time
behavior. The Fermi-Golden rule will build up as the time grows
beyond the inverse pair-excitation spectrum width, which renders
the quantum kinetics a Markovian process. The crossover towards
the Markovian behavior becomes observable when the width of the
pair-excitation spectrum is greater than or comparable to the
inverse of the initial dephasing time $T_{\rm H}^{\rm sh}$. For
small quantum dots, the hyperfine coupling and its gradient is
larger, resulting in a broader excitation spectrum, and therefore
the crossover behavior is observable, as shown in Fig.~\ref{size}.
For large quantum dots, where the pair-excitation spectrum is
relatively narrower, the initial stage decay could already
eliminate the spin coherence and thus the whole dephasing process
could be described by the $e^{-(2\tau /T_{\rm H}^{\rm sh})^4}$
profile.

The decoherence due to the intrinsic nuclear interaction is
suppressed by the echo pulse as evidenced by the enhancement of
the short-time decoherence time, $T_{\rm H}^{\rm sh}=\sqrt{2}
T_{2,B}$ (see Fig.~\ref{FID_sep12} and Fig.~\ref{times}). The
onset of the Markovian crossover also manifests itself in the
difference between the initial echo time $T_{\rm H}^{\rm sh}$ and
the overall echo decay time $T_{\rm H}$ for small dots, as shown
in Fig.~\ref{times}.

\section{Summary} \label{sec-summary}

In conclusion, we have presented the quantum theory for the
electron spin decoherence by a bath of interacting nuclear spins
under strong magnetic field and at finite temperature.
Entanglement between electron and nuclear spins, established by
their coupled evolution, leads to the loss of electron spin
coherence. The solution to the electron spin coherence amounts to
solving the many-body dynamics of the interacting nuclear spin
bath, which are conditioned on different electron spin states. In
the timescale of interest, the nuclear bath dynamics is dominated
by pair-correlations among nuclear spins, which can be mapped into
independent pseudo-spin excitations. Within the pair-correlation
approximation, the electron-nuclear spin dynamics reduces to the
coupled evolution of the electron spin with the non-interacting
pseudo-spin excitations, for which exact solutions are found.

Decoherence behaviors in GaAs quantum dots are calculated as
examples. We have demonstrated the significance of the extrinsic
nuclear coupling mediated by the virtual electron spin flips,
which manifests itself in the strong field dependence of the FID
in single-system dynamics. The calculated electron spin
decoherence time in single-system FID varies from $\sim
0.1$~$\mu$s to $\sim 10$~$\mu$s for field strength from $1$~T to
$20$~T, and saturates as the hyperfine mediated coupling is
suppressed by stronger field. The spin-echo pulse not only
recovers the coherence lost by inhomogeneous broadening but also
eliminates the decoherence due to the hyperfine-mediated nuclear
pair-flips and reduces the decoherence by the intrinsic nuclear
interaction, leading to a spin echo decay time $\sim 10$~$\mu$s
independent of field strength in ensemble dynamics.

The theory presented here can be applied to other electron-nuclear
spin systems, e.g., donor impurities in silicon, and may also be
extended to even more general cases of a quantum object in contact
with an interacting bath.

\acknowledgments
This work was supported by  NSF DMR-0403465, ARO/LPS, and
DARPA/AFOSR.

\begin{appendix}

\section{The Hamiltonian}  \label{app-H}

The total Hamiltonian of the system of an electron and many nuclear spins is given  by
\begin{equation}
\hat{H} =  \hat{H}_{\rm e} + \hat{H}_{\rm N} + \hat{H}_{\rm eN} +
\hat{H}_{\rm NN},
\end{equation}
composed of the single spin Zeeman energies $\hat{H}_{\rm e} =
\Omega_e \hat{S}^z_e$ and $\hat{H}_{\rm N} =
\sum_{n}\omega_{n}\hat{J}^z_{n}$ in the applied magnetic field
along the $z$-axis, the hyperfine interaction $\hat{H}_{\rm eN}$
and the intrinsic nuclear-nuclear interaction $\hat{H}_{\rm NN}$.

The hyperfine interaction between the electron and the nuclear
spins consists of the isotropic Fermi contact interaction and the
anisotropic dipole-dipole interaction. The latter is negligible
since the electron wave function in a quantum dot is dominated by
the $s$-orbit states. The contact hyperfine interaction is given
by,\cite{Slichter}
\begin{subequations}
\begin{eqnarray}
\hat{H}_{\rm eN} & = &\sum_{n}  a_{n}\hat{{\mathbf S}}_e\cdot \hat{\mathbf J}_{n}, \\
a_{n }& = & \frac{\mu _{0}}{4\pi } \gamma_e
\gamma_{n} \frac{8\pi }{3}\left| \Psi \left(
{\mathbf R}_{n }\right)\right|^{2},
\label{Eq_hf_contact}
\end{eqnarray}
\end{subequations}
where $\mu_0$ is the vacuum magnetic permeability, ${\mathbf R}_{n
}$ denotes the coordinates of the $n$th nucleus, $\gamma_n$ is the
nuclear gyromagnetic ratio, and $\gamma_e$ is the electron
gyromagnetic ratio. It should be noted that while the effective
$g$-factor in the quantum dot determines the Zeeman energy
$\Omega_e$, the free electron $g$-factor +2.0023 should be used
for the hyperfine coupling. \cite{Yafet}

The intrinsic interaction between nuclear spins includes the
direct dipole-dipole interaction, the indirect interactions
mediated by virtual excitation of electron-hole pairs, and the
intra-nuclear quadrupole interaction. The direct dipole term is
given by \cite{Slichter}
\begin{eqnarray}
\hat{H}^d_{\rm NN} & = &  \sum_{n < m} \frac{\mu _{0}}{4\pi }
\frac{\gamma_{n}\gamma_{m}} {R^3_{n;m}} \Big(\hat{\mathbf
J}_{n}\cdot\hat{\mathbf J}_{m} \nonumber -\frac{3\hat{\mathbf
J}_{n}\cdot{\mathbf R}_{n;m}
  {\mathbf R}_{n;m}\cdot\hat{\mathbf J}_{m}}
{R^2_{n;m}}\Big) , \label{Eq_NN_dipolar} \\
\end{eqnarray}
with ${\mathbf R}_{n;m}\equiv{\mathbf R}_{n}-{\mathbf R}_{m}$. The
indirect nuclear interaction is mediated via the virtual
excitation of electron-hole pairs by the hyperfine interaction
between nuclei and valence
electrons.\cite{indirect_exchange_Bloembergen,indirect_exchange_Anderson,
indirect_exchange_Shulman1,indirect_exchange_Shulman3,indirect_exchange_Sundfors}
When the virtual excitation is caused by the Fermi-contact
hyperfine interaction, the indirect interaction has the isotropic
exchange form
\begin{eqnarray}
\hat{H}^{\rm  ex}_{\rm NN}=-\sum_{n < m} B^{\rm  ex}_{n;m}\hat{\mathbf J}_{n}\cdot\hat{\mathbf J}_{m},
\label{Eq_NN_exchange}
\end{eqnarray}
named the pseudo-exchange interaction in the literature. The
leading contribution of the pseudo-exchange interaction for
nearest neighbors in the host crystal can be expressed
as\cite{indirect_exchange_Anderson,indirect_exchange_Shulman1}
\begin{eqnarray}
B^{\rm  ex}_{n;m} = \frac{\mu _{0}}{4\pi } \frac{ \gamma^{\rm
ex}_n \gamma^{\rm ex}_m}{ {R^{3}_{n;m}}} \frac{a_0}{R_{n;m}},
\label{Eq_NN_exchange_tensor}
\end{eqnarray}
where $\gamma^{\rm ex}_n$ is the effective gyromagnetic ratio
determined by the renormalized charge density of the $s$-orbit
electron. The indirect exchange interaction has been
experimentally studied by many researchers
\cite{indirect_exchange_Shulman1,indirect_exchange_Shulman3,indirect_exchange_Sundfors}.

When the virtual excitation of electron-hole pairs involves both
the Fermi-contact and the dipolar hyperfine interactions, group
theoretical analysis shows that the indirect nuclear spin
interaction has the dipolar form and is denoted as indirect
pseudo-dipolar interaction in the
literature.\cite{indirect_exchange_Bloembergen} Lattice distortion
can result in local electric field gradients, inducing the
intra-nuclear quadrupole interaction for nuclear spins with moment
greater than $1/2$. The indirect pseudo-dipolar interaction and
the quadrupole interaction are not included in the numerical
calculation in this paper due to the lack of experimental
characterizations in the literature. Nonetheless, they can be well
incorporated in our theory as contributions to energy cost and
transition amplitude for nuclear pair-flips [see
Eqs.~(\ref{Eq_Bk}) and (\ref{Eq_Dk})] when reliable data is
available. Furthermore, available studies
\cite{indirect_exchange_Shulman1,indirect_exchange_Shulman3,indirect_exchange_Sundfors}
show that the direct dipolar interaction together with indirect
exchange interaction can mainly account for the broadening and
lineshapes of NMR and NAR signals in the semiconductor matrix of
our interest and thus shall be the main ingredients of the
intrinsic nuclear interactions.

\section{The extrinsic Hyperfine-mediated nuclear spin-spin interaction}
\label{Append_transformation}

The contact hyperfine interaction can be separated into the
longitudinal (or diagonal) part
\begin{eqnarray}
\hat{H}_{\rm{eN},\it{l}}\equiv \sum_{n} a_{n}\hat{S}^z_e
\hat{J}^z_{n},
\end{eqnarray}
and the transverse (or off-diagonal) part
\begin{eqnarray}
\hat{H}_{\rm{eN},\it{t}}\equiv \sum_{n} \frac{a_{n}}{2}
\left(\hat{S}^+_e \hat{J}^-_{n}+\hat{S}^-_e \hat{J}^+_{n}\right).
\end{eqnarray}
The off-diagonal hyperfine interaction in the lowest order is eliminated by
the standard canonical transformation,
\begin{eqnarray}
\hat{W} \equiv \exp\left[\sum_{n}
\frac{a_{n}}{2\left(\Omega_e-\omega_n\right)} \left(\hat{S}^+_e
\hat{J}^-_{n}-\hat{S}^-_e \hat{J}^+_{n}\right)\right].
\end{eqnarray}
The residual second order term in the transformed Hamiltonian
$\hat{H}_{\rm red} = \hat{W} \hat{H} \hat{W}^{-1}$ is $\hat{H}_{A}
$ in Eq.~(\ref{HA}). We neglect higher order terms whose effects
are by at least a factor of $\sim (\Omega_e
\sqrt{N}/\mathcal{A})^{-2} \ll 1$ smaller as compared to
$\hat{H}_{A} $. Note that while the zeeman energy $\Omega_e \sim
10 - 100 ~\mu$eV for field $\sim 1-10~$T,\cite{Gurudev}
$\mathcal{A}/\sqrt{N} \sim T_2^{\ast -1} \sim 0.1~\mu$eV in GaAs
fluctuation dots.\cite{Gammon_T2star}

The canonical transformation also rotates the basis states, $|\pm
\rangle \otimes | \mathcal{J} \rangle$, $|{\mathcal J}\rangle
\equiv \bigotimes_{n}|j_{n}\rangle$, by a small amount
\begin{eqnarray}
\hat{W} |\pm\rangle\otimes|{\mathcal J}\rangle &\approx&
\left(1-\frac{1}{2}\sum_{n} \left|w^{\pm}_{n}\right|^2\right)
|\pm\rangle\otimes |{\mathcal J}\rangle \notag \\
&\mp& \sum_{n}w^{\pm}_{n}|\mp\rangle \otimes |j_{n}\pm 1\rangle
\bigotimes_{m\ne n}|j_{m}\rangle,
\end{eqnarray}
where we have kept only the lowest order correction in terms of
the expansion coefficient
\begin{equation}
w^{\pm}_{n} \equiv \frac{a_{n}}{2\left(\Omega_e-\omega_n\right)}
\sqrt{j \left(j +1\right)-j_{n}\left(j_{n}\pm 1\right)},
\end{equation}
which is small number due to the inequalities, $\Omega_e \gg
\omega_{n} \gg a_n$. $j=3/2$ for all three relevant isotopes
$^{75}\rm{As}$, $^{69}\rm{Ga}$ and $^{71}\rm{Ga}$.

\begin{widetext}
The rotation of the state vector actually accounts for the rapid
initial drop of the electron spin coherence. We show below that
this reduction in the visibility of the spin beat is negligible
under the field strength $\gtrsim 1~$T. The state evolution in the
very initial stage ($t\lesssim a_{n}^{-1}$) can be expressed as
\begin{eqnarray}
e^{-i\hat{H}t}|\pm\rangle\otimes|{\mathcal J}\rangle &=&
\hat{W}^{-1} e^{-i\hat{H}_{\rm red }t} \hat{W}
|\pm\rangle\otimes|{\mathcal J}\rangle \approx \hat{W}^{-1}
e^{-i\left(\hat{H}_{\rm e} + \hat{H}_{\rm N}
+ \hat{H}_{\rm{eN},\it{l}}\right)t}\hat{W} |\pm\rangle\otimes |{\mathcal J}\rangle \nonumber \\
&\approx & e^{\mp i\frac{1}{2}(\Omega_e + \mathcal{E}_{\mathcal
J}) t} e^{- i \sum_n j_n \omega_n t} \Bigg \{ \Big [
1-\sum_{n}\left(1-e^{\pm i\left(\Omega_e + \mathcal{E}_{\mathcal
J}-\omega_n \pm a_{n}\right)t}\right) \left|w^{\pm}_{n}\right|^2
\Big ] |\pm\rangle\otimes|{\mathcal J}\rangle
\nonumber \\
&& \pm \sum_{n} \left[1-e^{\pm i\left(\Omega_e +
\mathcal{E}_{\mathcal J}-\omega_n \pm a_{n}\right)t}\right]
{w}^{\pm}_{n} |\mp\rangle \otimes |j_{n}\pm 1\rangle
 \bigotimes_{m\ne n}|j_{m}\rangle \Bigg \},
\end{eqnarray}
\end{widetext}
where $\mathcal{E}_{\mathcal J}=\sum_{n}j_{n}a_{n}$ is the
Overhauser energy of the electron spin under the nuclear
configuration $| \mathcal{J} \rangle$ which arises from the
longitudinal (or diagonal) hyperfine interaction. To single out
the process of visibility loss, we have omitted above the
nuclear-nuclear coupling terms in $\hat{H}_{\rm
red}$.\cite{Loss_decoherence_nuclei} The reduced density matrix of
the electron spin can be derived as
\begin{subequations}
\begin{eqnarray}
\rho^{\rm e}_{+,+}(t) &\approx &  \rho^{\rm e}_{+,+}(0)
\left[1-p_+(t)\right]+\rho^{\rm e}_{-,-}(0)  p_-(t), \ \ \ \ \ \ \ \ \\
\left|\rho^{\rm e}_{+,-}(t)\right| &\approx & \left|\rho^{\rm
e}_{+,-}(0)\right|
      \left[1 -\frac{ p_+(t)}{2}   -\frac{ p_-(t)}{2}   \right],
\end{eqnarray}
\end{subequations}
with
\begin{eqnarray}
 p_{\pm}(t)  \equiv  4\sum_{n}\left({w}^{\pm}_{n}\right)^2
 \sin^2\frac{\left(\Omega_e + \mathcal{E}_{\mathcal J}-\omega_n\pm
 a_{n}\right)t}{2}. \label{visibilityloss}
\end{eqnarray}
The interference between different frequency components on the RHS
of Eq.~(\ref{visibilityloss}) will lead to a visibility loss of
the coherence and the initial drop of the population with
amplitude $\sim p_{\pm} \lesssim \left(\Omega_{\rm e} \sqrt{N} /
\mathcal{A} \right)^{-2}\ll 1$, occurring in the timescale $\sim
a_{n}^{-1}$. Therefore, in the high field limit ($\gtrsim 1~$T),
the electron spin flip by the nuclear spins is efficiently
suppressed ($T_1 \rightarrow \infty$ if phonon mechanisms
excluded).

\begin{figure}[t]
\includegraphics[width=6.3cm, height=6.5cm, bb=80 265 395 640,clip=true]{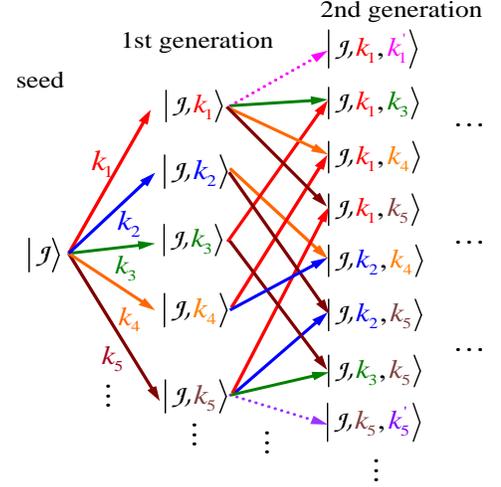}
\caption{(color online). The hierarchy structure for the nuclear
spin state evolution driven by pair-wise flip-lops. The $k'_1$ and
$k'_5$ pair-flips are possible only when the pair-flips $k_1$ and
$k_5$ have taken place, respectively. Some pair-flips occurs
exclusively to each other, such as $k_2$ and $k_3$.}
\label{Fig_hierarchy}
\end{figure}

To conclude, the transverse (or off-diagonal) part of the
hyperfine interaction has two effects: (i) the transformation
acting on the Hamiltonian results in an effective coupling between
the nuclear spins ($\hat{H}_A$ in Eq.~(\ref{HA})), which
contribute to the pure dephasing of the electron spin coherence;
(ii) the transformation acting on the state vector can be
understood as a visibility loss. While the two processes coexist,
we have shown that under the field strength $\gtrsim 1~$T, the
visibility loss is negligibly small. Therefore, the exact
evolution of the electron nuclear system can be well approximated
as
\begin{equation}
e^{-i\hat{H}t}|\pm\rangle\otimes|{\mathcal J} \rangle \backsimeq
e^{-i\hat{H}_{\rm red}t}|\pm\rangle\otimes|{\mathcal J} \rangle
\end{equation}
with the effect of the transverse hyperfine interaction well
incorporated as the extrinsic nuclear coupling $\hat{H}_A$.

\section{Solution of the dynamics of the electron-nuclear spin system} \label{A_solution}

The elementary process driven by the interaction Hamiltonian,
Eq.~(\ref{eq-Hpm}), is the pair-wise homo-nuclear spin flip-flop.
Here, we examine the dynamics of the basic flip-flop process,
based on which we build a hierarchical framework of many-spin
basis states for dynamics, investigate the spin correlations, and
construct a pseudo-spin method.

\subsection{The basic nuclear spin excitation}

The transition for the pair-flip driven by the operator
$\hat{J}^+_{n}\hat{J}^-_{m}$ consists in a pair of nuclear spins
in state $|j_n,j_m\rangle$ goes to $|j_n+1,j_m-1\rangle$ if
permitted, i.e., $j_n+1 \le j$ and $j_m-1\ge - j$ for spin $j$.
The shorthand $k$ is used to denote this pair transition. Such a
$k$-th pair-flip transition between two many-nuclear spin states
is described as
\begin{eqnarray}
|\pm\rangle|{\mathcal J}\rangle\longrightarrow |\pm\rangle|{\mathcal J},k\rangle.
\label{Eq_J_flip}
\end{eqnarray}
The transition matrix element $\pm A_k+B_k$ and the energy cost $
D_k\pm E_k$, the sign conditioned on the electron spin state $|
\pm \rangle$, can be derived from Eqs.(\ref{Hamiltonian}),
\begin{widetext}
\begin{subequations}
\begin{eqnarray}
A_k &\equiv &  \left\langle {\mathcal J},k\right| \hat{H}_A \left|{\mathcal J}\right\rangle
 =  \frac{a_{n}a_{m}}{4\Omega_e}
\sqrt{j\left(j+1\right)-j_{n}\left(j_{n}+1\right)}
\sqrt{j \left(j +1\right)-j_{m}\left(j_{m}-1\right)}, \label{Eq_Ak}\\
B_k &\equiv &  \left\langle {\mathcal J},k\right| \hat{H}_B
\left|{\mathcal J}\right\rangle =B_{n,m} \sqrt{j \left(j + 1
\right)-j_{n}\left(j_{n}+1\right)}
\sqrt{j \left(j +1\right)-j_{m}\left(j_{m}-1\right)}, \label{Eq_Bk} \\
D_k &\equiv &  \left\langle {\mathcal J},k\right| \hat{H}_D
\left|{\mathcal J},k\right\rangle
 -  \left\langle {\mathcal J}\right| \hat{H}_D \left|{\mathcal J}\right\rangle =
\sum_{n'}D_{n,n'}j_{n'}-\sum_{m'}D_{m,m'}j_{m'} - D_{n,m}, \ \ \
\label{Eq_Dk}\\
E_k & \equiv &  \left\langle {\mathcal J},k\right| \hat{H}_E \left|{\mathcal J},k\right\rangle
 -  \left\langle {\mathcal J}\right| \hat{H}_E \left|{\mathcal J}\right\rangle
=\left(a_{n}-a_{m}\right)/2. \label{Eq_Ek}
\end{eqnarray}
\label{elementaryexcitation}
\end{subequations}
\end{widetext}

\subsection{The hierarchy of nuclear pair-flip states}

The evolution of the nuclear spin state $|{\mathcal
J}^{\pm}(t)\rangle \equiv e^{-i \hat{H}^{\pm} t} | \mathcal{J}
\rangle $ can be formally described by pair-flip transitions in a
hierarchy of basis states. The hierarchy is composed of the seed
state, $|{\mathcal J}\rangle$, the first-generation states, each
$|{\mathcal J}, k\rangle$ generated from the seed state by the
$k$th pair-flip, the second-generation states, each  $|{\mathcal
J}, k_1, k_2\rangle$ generated from the first-generation state
$|{\mathcal J}, k_1\rangle$ or $|{\mathcal J}, k_2\rangle$  by the
$k_1$th or the $k_2$th pair-flip, respectively, and so on. A
many-nuclear spin state can be expanded in this basis as
\begin{eqnarray}
|{\mathcal J}^{\pm}(t)\rangle  &= &
C^{\pm}_{\mathcal J}(t)|{\mathcal J}\rangle
+\sum_{k_1} C^{\pm}_{{\mathcal J},k_1}(t)|{\mathcal J},k_1\rangle \nonumber \\ &&
+\sum_{k_1,k_2} C^{\pm}_{{\mathcal J},k_1,k_2} (t) |{\mathcal J},k_1,k_2\rangle+\cdots, \ \ \
\label{expansion}
\end{eqnarray}
in which the wavefunction satisfies the equation
\begin{eqnarray}
&& \partial_tC^{\pm}_{{\mathcal J},k_1,\ldots,k_p}
=
-iE^{\pm}_{{\mathcal J},k_1,\ldots,k_p}C^{\pm}_{{\mathcal J},k_1,\dots,k_p}
\nonumber \\ && \phantom{\partial_tC^{\pm}}
-i\sum_{j=1}^p\sum_{k_{j}}\left(B_{k_j}\pm A_{k_j}\right)C^{\pm}_{{\mathcal J},k_1,\ldots,k_{j-1},k_{j+1},\ldots,k_{p}}
\nonumber \\ && \phantom{\partial_tC^{\pm}}
-i\sum_{k\ne k_1,\ldots,k_p}\left(B^*_{k}\pm A^*_{k}\right)C^{\pm}_{{\mathcal J},k_1,\ldots,k_p,k},
\label{Schrodinger}
\end{eqnarray}
where the energy $E^{\pm}_{{\mathcal J},k_1,\ldots,k_p}$ is the
eigenenergy of the basis state $|{\mathcal
J},k_1,\ldots,k_p\rangle$ under the Hamiltonian $\hat{H}_D\pm
\hat{H}_E$, respectively. The hierarchy description of the nuclear
spin dynamics is illustrated in Fig.~\ref{Fig_hierarchy}.

\begin{figure}[t]
\includegraphics[width=6.1cm, height=4.5cm, bb=75 505 380 730,clip=true]{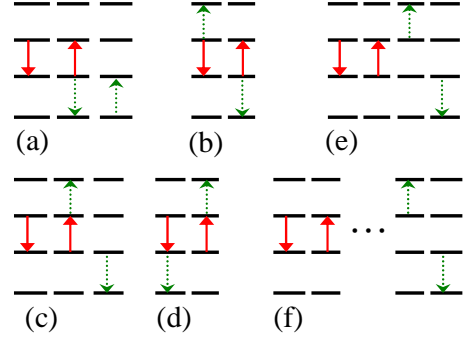}
\caption{(Color online). Possible correlations between two nuclear
spin pair-flips ($k_1$ and $k_2$, indicated by solid and dotted
arrows, respectively). (a) and (b), the two pair-flips sharing one
or two nuclei occur exclusively to each other. (c) and (d), one
pair-flip can take place only after the other pair has been
flipped. (e) The energy cost $D_k$ of one pair-flip depends on
whether or not the other pair has been flipped, when they involve
spins in neighborhood. (f) The two pair-flips are independent of
each other when they are far apart from each other.}
\label{Fig_correlation}
\end{figure}

The many-body nature of the problem lies in the fact that the
pair-wise flip-flops are correlated in general. The correlation
between two pair-flips can be developed in the following three
cases:
\begin{enumerate}
\item {\em Exclusive pair-flips} as shown in Fig.~\ref{Fig_correlation} (a) and (b).
 When one pair-flip ($k_1$) has taken place, the other one ($k_2$) which shares one or two
 spins with the flipped pair cannot occur any more, and vice versa.
\item {\em Subsequent pair-flips} as shown in Fig.~\ref{Fig_correlation} (c) and (d).
 When two pair-flips share one or two nuclear spins, one pair-flip is possible only after
 the other one has already taken place.
\item {\em Neighboring pairs} as shown in Fig.~\ref{Fig_correlation} (e).
 Two pair-flips can be correlated even when they don't share a nuclear spin but
 are in neighborhood, since one pair-flip will change the nuclear spin configuration in the neighborhood
 of the other pair and thus modify the energy cost ($D_k$) due to the diagonal nuclear spin interactions of one
 pair-flip. This correlation can be clearly seen from Eq.~(\ref{Eq_Dk}).
\end{enumerate}
When two pair-wise flip-flops are not in neighborhood [Fig.~\ref{Fig_correlation} (f)],
they are independent of each other.

With increasing numbers of pair-flips, the nuclear spin state
involves basis states further up in the hierarchy structure
depicted in Fig.~\ref{Fig_hierarchy}, and higher order
correlations may occur, making the solution of nuclear spin
dynamics a formidable task in general.

\subsection{Pair-correlation approximation (PCA)} \label{SS_PCA}

Two pair-flips are correlated only when they are located in
neighborhood. We can estimate the probability of having pair-flips
correlated in the $p$th generation of the hierarchy: if $p-1$
pair-flip excitations have been generated, the probability of
having the $p$th pair outside the neighborhood of all the previous
ones is about $(1-q(p-1)/N)$. By induction, the probability of
having $p$ pair-flips uncorrelated is
\begin{eqnarray}
1- P_{\rm corr}\sim \prod_{j=1}^{p-1}\left(1-qj/N\right)\sim
\exp\left(-qp^2/N\right). \label{ProbabilityUncorrelated}
\end{eqnarray}

The number of pair-flip excitations at time $t$ may be estimated
by
\begin{eqnarray}
N_{\rm flip}\equiv \max_{\mu=\pm} \left(\sum_{p=1}^{\infty}
p\left|C^{\mu}_{J,k_1,\ldots,k_p}\right|^2\right) \label{Npair}
\end{eqnarray}
If $q N_{\rm flip}^2(t)\ll N$ in the timescale of interest, the
probability, $P_{\rm corr}$, of having pair-flip excitations
correlated is negligibly small. Thus, by removing the few states
reached via subsequently correlated pair-flips and adding few
states containing exclusive pair-flips, the exact Hilbert space
can be mapped into the tensor product of two-dimensional Hilbert
subspaces, each of them corresponding to a pair-flip available
from the seed state $|{\mathcal J}\rangle$, namely
\begin{eqnarray}
\left\{|{\mathcal J}\rangle, |{\mathcal J}, k_1\rangle, |{\mathcal
J},k_1,k_2\rangle,\ldots \right\} \longrightarrow \bigotimes_k
\left\{|\uparrow_k \rangle, |\downarrow_k \rangle\right\},
\end{eqnarray}
where the index $k$ runs over all possible pair-flips from the
seed state to the first generation. The mapping can be explicitly
expressed as
\begin{subequations}
\begin{eqnarray}
|{\mathcal J}\rangle & \longrightarrow & \bigotimes_k |\uparrow_k \rangle, \\
|{\mathcal J}, k_1 \rangle & \longrightarrow  & |\downarrow_{k_1} \rangle \bigotimes_{k\ne k_1} |\uparrow_k \rangle, \\
|{\mathcal J}, k_1,k_2\rangle & \longrightarrow & |\downarrow_{k_1} \rangle |\downarrow_{k_2}\rangle
\bigotimes_{k\ne k_1,k_2} |\uparrow_k \rangle, \\
\cdots  & \cdots & \cdots  \nonumber
\end{eqnarray}
\label{mapping}
\end{subequations}
and the nuclear bath state can be factorized into elementary
excitations as
\begin{eqnarray}
\left|{\mathcal J}^{\pm}(t)\right\rangle = e^{-iE^{\pm}_{\mathcal
J}t}\bigotimes_k \left(g_{k}^{\pm}(t)| \uparrow_k \rangle +
f_{k}^{\pm}(t)|\downarrow_k \rangle \right). \label{PCAstate}
\label{Uncorrelatedwavefunction}
\end{eqnarray}
Furthermore, the energy cost of a pair-flip is assumed independent
of whether or not another pair-flip has occurred in the
neighborhood. So all the elementary excitations are treated
independent of each other and their dynamics is determined solely
by their own energy costs and transition matrix elements as
\begin{subequations}
\begin{eqnarray}
i\partial_t g_{k}^{\pm}&=&\left(B_k\pm A_k\right)f_{k}^{\pm}, \\
i\partial_t f_{k}^{\pm}&=&\left(D_k\pm E_k\right)f_{k}^{\pm} +
\left(B_k\pm A_k\right) g_{k}^{\pm}, \ \ \ \ \ \ \
\end{eqnarray}
\label{elementarydynamics}
\end{subequations}
with initial conditions $g_{k}^{\pm}(0)=1$ and $f_{k}^{\pm}(0)=0$.
The above approximation strategy is denoted as the
pair-correlation approximation (PCA) where all pair-correlations
among the nuclear spins are kept and higher order nuclear
correlations are neglected.

In PCA, we will calculate physical properties by using
Eq.~(\ref{PCAstate}) as the bath state at time $t$ instead of the
exact state Eq.~(\ref{Schrodinger}). In this paper, the electron
spin coherence of interest is essentially the state overlap
between two differently driven bath state as shown in
Eq.~(\ref{eq-losch}). The relative amount of change to the bath
Hilbert space structure by PCA is $P_{\rm corr}$. Therefore, the
relative error of PCA in calculating the coherence is bounded by
\begin{eqnarray}
\epsilon \lesssim P_{\rm corr} \sim 1-\exp\left(-q N^2_{\rm
flip}/N\right). \label{error}
\end{eqnarray}

\subsection{The pseudo-spin model} \label{Asub_PseudoSpin}

The uncorrelated pair-flip dynamics given in
Eq.~(\ref{elementarydynamics}) is nothing but the independent
evolutions of two-level systems $\{ |\uparrow_k\rangle,
|\downarrow_k \rangle \}$. Thus the electron-nuclear spin dynamics
$| \mathcal{J}^{\pm} (t)\rangle$ from the initial state $|
\mathcal{J} \rangle$ is mapped, by PCA, to the precession of
${\mathcal N}$ non-interacting pseudo-spins under pseudo-fields
conditioned on the electron spin state
\begin{subequations}
\begin{eqnarray}
\hat{\mathcal H}_{k}^{\pm} &\equiv& {\bf h}_{k}^{\pm}\cdot \hat{\boldsymbol \sigma}_k /2,\\
{\bf h}_{k}^{\pm} &\equiv & \left(\pm 2A_k+2B_k,0,D_k\pm
E_k\right), \label{pseudofield}
\end{eqnarray}
\end{subequations}
where $\hat {\boldsymbol \sigma}_k\equiv
\left(\hat{\sigma}^x_k,\hat{\sigma}^y_k,\hat{\sigma}^z_k\right)$
is the Pauli matrix for the pseudo-spin corresponding to the $k$th
pair-flip available from the seed state $|{\mathcal J}\rangle$.
Note that by the Hamiltonian mapping from Eq.~(\ref{eq-Hpm}) to
Eq.~(\ref{pseudospin_Hamil}), the energy of the seed state has
been shifted by a constant $\sum_k\left(D_k\pm E_k\right)/2$,
which does not affect the calculation of $| \langle
\mathcal{J}^-(t) | \mathcal{J}^+ (t) \rangle |$. The directions
(indicated by the superscripts $x$, $y$, and $z$) for the
pseudo-spins are not defined in the real-space but in a fictitious
pseudo-space.

\section{Range of validity of the pair-correlation approximation} \label{A_validity}

Within the PCA, the averaged pair-flip number $N_{\rm flip}$
defined in Eq.~(\ref{Npair}) has a simpler form
\begin{equation}
N_{\rm flip}= \max_{\pm}\left(\sum_k |\langle \uparrow | e^{-i
\hat{\mathcal H}^{\pm}_k t} | \uparrow \rangle|^2 \right),
\label{Npair_PCA}
\end{equation}
with which we can make an {\it a posterior} self-consistency check
of the validity of the approximation. If $N_{\rm flip}$ obtained
from Eq.~(\ref{Npair_PCA}) is small, we conclude that it also
faithfully reflects the number of pair-flip excitations in the
exact dynamics and therefore, the error estimation would be
faithful and the physical property can be calculated based on PCA
with a relative error bounded by Eq.~(\ref{error}).

The total number of pair-flip excitations can be divide into two
parts: $N_{\rm flip}\left( t\right) =N_{\rm flip,A}\left( t\right)
+N_{\rm flip,B}\left( t\right) $, where $N_{\rm flip,A}$ is the
number of non-local pair-flip excitations and $N_{\rm flip,B}$ is
the number of local pair-flip excitations that have been created.
$ N_{\rm flip,A}\left( t\right) $ and $N_{\rm flip,B}\left(
t\right) $ have very different dependence on time and system
parameters, and we analyze them separately.

In free-induction evolution, the number of non-local pair-flip
excitations is given by,
\begin{eqnarray}
N_{\rm flip,A}\left( t\right)& = &\sum_{k}\left( \frac{
2A_{k}}{\sqrt{ E_{k}^{2}+4A_{k}^{2}}}\right) ^{2}\sin
^{2}\frac{\sqrt{ E_{k}^{2}+4A_{k}^{2}} t }{2}
\notag \\
&\leq& \sum_{k} A_{k}^{2} t^2 \sim \mathcal{N}_A
\frac{\mathcal{A}^4}{N^4\Omega_e^2} t^2 \sim
\frac{\mathcal{A}^4}{N^2\Omega_e^2} t^2
\end{eqnarray}
where $\mathcal{N}_A \sim N^2$ is the number of non-local nuclear
spin pairs. Since the evolution of the non-local pair-correlation
is completely reversed by the $\pi$ pulses (see the discussion in
Section \ref{sec-results}), $N_{\rm flip,A}\left( t\right)$ is
also reversed and returns to zero at each spin echo time.
Therefore, $N_{\rm flip,A}\left( t\right) $ does not accumulate in
the pulse controlled dynamics \cite{yls2} and we just need to look
at the maximum value of $N_{\rm flip,A}\left( t\right) $ between
echoes.

For single system FID and ensemble spin echo calculation presented
in this paper, the sufficient condition for PCA to be valid is
$N^2_{\rm flip,A}\left( T_H \right) \ll N$ by noticing that $T_2
\thickapprox \min(T_{2,A},T_{2,B})$ and $T_H =\sqrt{2} T_{2,B}$.
From Eq.~(\ref{size_field_depend}), we have,
\begin{eqnarray}
N_{\rm flip,A}\left( T_H \right) &\sim& N^{-3/2}\Omega_e^{-2}
b^{-1} \mathcal{A}^3,
\end{eqnarray}
Therefore, we obtain a lower bound on the dot size $N$ for the
validity of PCA: $N^4 \gg \mathcal{A}^6 \Omega_e^{-4} b^{-2}$. For
GaAs fluctuation dot in a field of $10~$T, the above condition is
well satisfied for $N \gtrsim 10^4$.

In contrast to the non-local pair dynamics, the local pair
dynamics is {\it not} reversed under the influence of the electron
spin flip and $N_{\rm flip,B}(t)$ accumulates all through the
time. Nonetheless, it turns out that for all scenarios of interest
including the evolution under the pulse sequence control as in
\cite{yls2}, we have,
\begin{eqnarray}
N_{\rm flip,B}\left( t\right) \leq \sum_{k} B_{k}^{2} t^2 \sim
\mathcal{N}_B b^2 t^2 \sim N b^2 t^2
\end{eqnarray}
where $\mathcal{N}_B \sim N$ is the number of local nuclear spin
pairs. For the relevance of the single system FID and ensemble
spin echo calculation, we shall examine
\begin{eqnarray}
N_{\rm flip,B}\left( T_{H} \right) &\thickapprox& N^{3/2} b
\mathcal{A}^{-1}
\end{eqnarray}
and therefore, the condition $(N_{\rm flip,B})^2 \ll N$ sets an
upper bound on the quantum dot size $N$: $N^2  b^2
\mathcal{A}^{-2} \ll 1$. In GaAs, the above condition is well
satisfied for quantum dot size $N \lesssim 10^8$.

To conclude, for nuclear pair-correlations to dominate over higher
order correlations in the relevant timescale, local nuclear
pair-dynamics driven by the intrinsic couplings imposes an upper
bound on $N$ while non-local nuclear pair-dynamics driven by the
extrinsic coupling imposes a lower bound. This mesoscopic regime
covers quantum dots of all practical size. Within this mesoscopic
regime, higher order nuclear correlations are well negligible
under the timescale of interest. The error estimation is based on
characterizing the difference in the Hilbert space structure of
the exact dynamics and that of the pseudo-spin model and assuming
this difference has a full influence on the electron spin
coherence calculation. Therefore, the bound of Eq.~(\ref{error})
is not necessarily a tight bound.

\end{appendix}

\end{document}